\documentclass[12pt]{article}
\usepackage{amsmath}
\usepackage{graphicx,psfrag,epsf}
\usepackage{enumerate}
\usepackage{natbib}
\usepackage{comment}
\usepackage{xcolor}
\usepackage{authblk}
\usepackage{url} 

\def\bSig\mathbf{\Sigma}

\begin{document}

\def\spacingset#1{\renewcommand{\baselinestretch}%
{#1}\small\normalsize} \spacingset{1}


\title{\bf Doubly Robust Estimation of Desirability of Outcome Ranking (DOOR) Probability with Application to MDRO Studies}
\author[1,2]{Shiyu Shu}
\author[1,2]{Toshimitsu Hamasaki}
\author[1,2]{Scott Evans}
\author[2]{Lauren Komarow}
\author[3]{David van Duin}
\author[1,2]{Guoqing Diao\thanks{
Contact gdiao@email.gwu.edu
}}
\affil[1]{Department of Biostatistics and Bioinformatics, The George Washington University}
\affil[2]{The Biostatistics Center, The George Washington University}
\affil[3]{Division of Infectious Diseases, Department of Medicine, University of North Carolina at Chapel Hill}
\date{}
\maketitle

\begin{abstract}
In observational studies, adjusting for confounders is required if a treatment comparison is planned. A crude comparison of the primary endpoint without covariate adjustment will suffer from biases, and the addition of regression models could improve precision by incorporating imbalanced covariates, and thus help make correct inference. Desirability of outcome ranking (DOOR) is a patient-centric benefit-risk evaluation methodology designed for randomized clinical trials. Still, robust covariate adjustment methods could further expand the compatibility of this method in observational studies. In DOOR analysis, each participant’s outcome is ranked based on pre-specified clinical criteria, where the most desirable rank represents a good outcome with no side effects and the least desirable rank is the worst possible clinical outcome. We develop a causal framework for estimating the population-level DOOR probability, via the inverse probability of treatment weighting method, G-Computation method,  and a Doubly Robust method that combines both. The performance of the proposed methodologies is examined through simulations. We also perform a causal analysis of the Multi-Drug Resistant Organism (MDRO) network within the Antibacterial Resistant Leadership Group (ARLG),  comparing the benefit:risk between Mono-drug therapy and Combination-drug therapy.
\end{abstract}

\noindent%
\vfill

\spacingset{1.45} 

\newpage

\section{Introduction}

Randomization plus the Intention-to-Treat (ITT) principle is the gold standard for biomedical studies, as randomization enables researchers to uncover the causal effect of an intervention. However, we do not always have the luxury of conducting a randomized study and instead we need to rely on observational studies to answer research questions; for example, post-marketing surveillance using registry-based data can detect long-term effects and safety signals \citep{c2}; Post-vaccine monitoring for side effects in rare populations such as pregnant women or patients with autoimmune diseases can help guide policy-making \citep{c3}; Evaluation of health outcomes after the initiation of national programs such as the National Diabetes Prevention Program can offer insights into the success of the program \citep{c4}. In such scenarios, observational studies could either provide direct scientific evidence, or assist clinical development in providing real-world evidence (RWE, \cite{c35}), if the analysis is conducted appropriately. An imbalance among covariates/confounders could lead to biased crude estimation of the causal effects, and thus incorrect inference, which is a primary concern in observational studies. Causal inference techniques could mitigate bias by adjusting for confounders in statistical models. Popular options include Inverse Probability to Treatment Weighting (IPTW) \citep{c7}, G-Formula \citep{c8}, Augmented IPTW \citep{c9}, etc.\\

When conducting biomedical research either in an interventional setting or an observational setting, benefit:risk analysis is an important tool to increase understanding about a novel intervention. Traditional benefit:risk analysis evaluates efficacy and safety marginally. It thus fails to incorporate associations between outcomes or recognize the cumulative nature of outcomes in individual patients, and suffers from competing risk complexities during the interpretation of individual outcomes \citep{c22}. In the last 15 years, a family of nonparametric methods that utilize Wilcoxon-Mann-Whitney U statistics gained popularity. Namely, \cite{c10} proposed Generalized Pairwise Comparison to analyze Proportion in Favor of Treatment/Net Treatment Benefit/Somers' D; \cite{c11} proposed the Win Ratio method; and \cite{c12} proposed the Desirability of Outcome Ranking (DOOR) \citep{c36, c22} to estimate DOOR probability, all aiming to evaluate the treatment effect at the patient level, and use outcomes to analyze patients.\\

DOOR is a novel paradigm for the design, analysis, and interpretation of clinical trials that allows physicians to create better treatment strategies by giving them a way to review data and compare the risks and benefits of interventional options. The construction of a DOOR outcome requires tailoring to the disease of interest and research question of a clinical trial. For example, \cite{c13} proposed a DOOR outcome for the evaluation of antibacterial resistance treatment, which has 5 ranked categories and the rank is determined by the survival status and the total number of component events experienced by patients.\\

Once the ranking of $K$ different patient status is established, the estimand DOOR probability, which measures the probability that a patient in the treatment arm has a more desirable DOOR outcome than a patient in control, could be established. Specifically, suppose that there are $n_1$ participants in treatment A and $n_2$ participants in treatment B, $n_1n_2$ patient-pair comparisons can be formed in total. Among these pairwise comparisons, we assign value 1 if A is more desirable, 0.5 for the ties, and 0 if B is more desirable, sum up the value of each pairwise comparison, and divide the sum by $n_1n_2$. DOOR probability measures whether a randomly selected patient in A could have a more desirable outcome than a randomly selected patient in B at the population level, and DOOR probability equal to 0.5 could be the basis of hypothesis testing.\\

Most current DOOR analyses focus on clinical trial data, so no further covariate adjustment is necessary. Some propensity score weighting methods were discussed before\citep{c28, c29, c30} in observational studies within the Antibacterial Resistance Leadership Group (ARLG), but such methods would rely on the correct specification of the propensity model, and no asymptotic properties of this estimator were discussed. A motivating example originated from ARLG highlights the importance of exploring/expanding covariate adjustment methods within the DOOR methodology: when clinicians treat patients with carbapenem-resistant bacteria after collection of lab cultures, they can prescribe a single drug (mono-therapy) or a combination of drugs (combination-therapy). The use of some combinations of antibiotics is supported by mostly non-clinical evidence, and the clinical efficacy of such combinations remains largely unknown. Furthermore, treatment with more than one antibiotic increases the likelihood of adverse events, and may ultimately result in worse outcomes. The condition of the patient at the time of antibiotic prescription is likely to have an impact on the choice of using one vs. two or more antibiotics for initial treatment. For instance, patients who are more acutely ill due to infection may be more likely to receive multiple vs. a single antibiotic. Therefore, covariate adjustment is necessary to better understand the benefit:risk profiles of mono vs. combination therapy.\\

In this paper, we develop causal inference methods under the DOOR framework. As DOOR probability is a function of multinomial probabilities, we consider three different approaches to re-weight/re-estimate these probabilities: 1. A propensity model that estimates the probability of receiving treatments given covariates, which was done within the Antibacterial Resistance Leadership Group (ARLG) network; 2. An outcome model that estimates the cell probabilities given covariates; 3. A doubly robust method that combines the merits of both methods and is less susceptible to model misspecifications.\\

In section 2, we describe the proposed method in detail. Section 3 provides simulation results, and an application to an infectious disease dataset is presented. We conclude the paper with discussions in Section 4. All technical details of the theoretical results are deferred to the Appendix. Additional figures and tables, along with the R code can be found in the Web Appendix.

\section{Methods}

We define $Y$ as the ordinal DOOR outcome with $K$ levels, $Z$ as the binary treatment indicator that equals 1 for the treatment group, and 0 for the control group, $\mathbf{X}$ as a vector of potential confounders. Without loss of generality, we assume that a larger $Y$ represents a more desirable patient status. For each individual $i(i=1,\cdots,n)$, we observe $O_i = \{Y_i, Z_i, \mathbf{X_i}\}$.\\

We define DOOR probability as $D = P(Y^1 > Y^0) + P(Y^1=Y^0)/2$ or equivalently:
\begin{equation}D=\sum_{k=2}^KP(Y^1=k)\sum_{l=1}^{k-1}P(Y^0=l)+\frac{1}{2}\sum_{k=1}^KP(Y^1=k)P(Y^0=k), \end{equation}
where $Y^1, Y^0$ represent counterfactual DOOR outcomes under treatment and control, respectively, and this quantity can be estimated by plugging in the observed cell probabilities. \\

We further define $P(Y^1) = [P(Y^1=1), P(Y^1=2), \cdots,P(Y^1=K)]^T$ as the vector of counterfactual probabilities under treatment, and $P(Y^0)$ likewise. Equation (1) can be re-expressed as $D=P(Y^1)^TAP(Y^0)$, where $A$ is a lower triangular matrix that has 0.5 on the diagonal, and 1 as the off-diagonal non-zero elements. Throughout this paper, we make two standard assumptions in the causal inference literature: 1. the strong ignorability assumption\citep{c31}, $\{Y^1,Y^0\} \perp\!\!\!\perp Z \mid \mathbf{X}$, i.e., the potential outcomes are independent from treatment conditional on confounders; 2. the positivity assumption, $\delta\leq P(Z=1\mid\mathbf{X})\leq 1-\delta$ almost surely for some $\delta\in (0,1)$, i.e., the probability of receiving treatment is positive given combinations of confounders.\\

With the causal setup in mind, we describe three approaches for estimating the DOOR probability between two arms.

\subsection{IPTW/Propensity Score Method (PS)}

We fit a statistical model for the probability that a participant will receive the treatment of interest given the confounders, i.e., the propensity score defined by $P(Z_i=1|\mathbf{X}_i)$. We consider the logistic regression model 
\begin{equation*}\text{logit}P(Z_i=1|\mathbf{X}_i)=\beta^T\mathbf{X}_i, i=1,\cdots,n \end{equation*}
where $\beta$ is a vector of coefficients. Denoting $\hat \beta$ as the maximum likelihood estimator of $\beta$, we have $\pi_i=\pi_i(\mathbf{X}_i,\beta)\equiv\exp(\beta^T\mathbf{X}_i)/\{1+\exp(\beta^T\mathbf{X}_i)\}$, where we call $\pi_i$ the propensity score. We can then estimate the probability distribution for the potential DOOR outcome in treatment arms $Z=1$ and $Z=0$, respectively, by 
\begin{equation*}\hat p_{1k}=\hat P(Y^1=k)=\frac{1}{n}\sum_{i=1}^n\frac{Z_iI(Y_i=k)}{\hat\pi_i}, k=1,\cdots,K \end{equation*}
and
\begin{equation*}\hat p_{0k}=\hat P(Y^0=k)=\frac{1}{n}\sum_{i=1}^n\frac{(1-Z_i)I(Y_i=k)}{1-\hat\pi_i}, k=1,\cdots,K \end{equation*} where $I(\cdot)$ is the indicator function. These IPTW probabilities do not necessarily sum up to 1, and we can use normalized weights (the Hajek estimator, \cite{c39}) by dividing each cell probability by the sum. We can then estimate the DOOR probability by plugging in the IPTW estimators, $\hat D=\hat P(Y^1)^TA\hat P(Y^0)$. Note that the IPTW method requires a correct specification for the propensity score model.

\subsection{G-Formula/Potential Outcome Method (PO)}

We consider fitting a model for the conditional distribution of the DOOR outcome in each treatment arm given confounders. For example, we may consider a proportional odds model such as $\log(P(Y\leq k\mid\mathbf{X})/P(Y>k\mid\mathbf{X})) = \alpha_k + \mathbf{X}^T\gamma, k=1,\cdots,K-1$. The conditional probability $P(Y_i=k|Z=a, \mathbf{X}_i)=m_a(\mathbf{X}_i,\Theta)=m_{iak}, (a=0,1)$, where $\Theta$ contains the regression parameters (intercepts $\alpha_1,\cdots,\alpha_{K-1}$ and coefficients $\gamma$) that can be estimated via the maximum likelihood method. We can estimate the DOOR probability through G-computation
\begin{equation*}\hat P(Y^a=k)=\frac{1}{n}\sum_{i=1}^n\hat m_{iak}, k=1,\cdots,K.\end{equation*}
As in the IPTW method, we can plug in the estimated cell probabilities to estimate $\hat D$. Note that the performance of G-Formula relies on the correct specification of the DOOR outcome model, which is not necessarily the same as the propensity model.

\subsection{Doubly Robust Method (DR)}

Both the IPTW method and the G-estimation method make certain model assumptions. If the specific model assumption is violated, the IPTW or the G-Formula will have biased estimators of the DOOR probability. To overcome the limitations, we can consider the
doubly robust estimation method. Specifically, we combine the propensity model and the outcome model mentioned above:
\begin{equation*}\hat p_{1k}=\hat P(Y^1=k)=\frac{1}{n}\sum_{i=1}^n\left\{\frac{Z_iI(Y_i=k)}{\hat\pi_i}-\frac{Z_i-\hat\pi_i}{\hat\pi_i}\hat m_{i1k}\right\}, k=1,\cdots,K \end{equation*}
and
\begin{equation*}\hat p_{0k}=\hat P(Y^0=k)=\frac{1}{n}\sum_{i=1}^n\left\{\frac{(1-Z_i)I(Y_i=k)}{1-\hat\pi_i}+\frac{Z_i-\hat\pi_i}{1-\hat\pi_i}\hat m_{i0k}\right\}, k=1,\cdots,K \end{equation*}
This result is a direct extension of Equation 9 in \cite{c14} by replacing $Y_i$ in their notation with the indicator function. The doubly robust estimators will be consistent as long as one of the two models is correctly specified.

\subsection{Asymptotic Results}

We consider a unified approach by utilizing Influence Functions (IF) \citep{c15} to establish the asymptotic properties of the proposed estimators and the corresponding limiting variances. Detailed derivations and definition of certain terms are included in the Appendix, and we will only present the formula here.

\newtheorem{theorem}{Theorem}
\newtheorem{lemma}{Lemma}
\newtheorem{definition}{Definition}

\begin{theorem}[Asymptotic Normality of Influence-Function Estimators]
Let $O_i = (Y_i, Z_i, \mathbf{X}_i),i=1,\cdots,n$ be i.i.d.\ observations from a distribution $P_0$,
and let $\psi(P)=[p_{11},p_{12},\cdots,p_{1(k-1)},p_{01},\cdots,p_{0(k-1)}]$ be a real-valued causal parameter vector with true value
$\psi_0 = \psi(P_0)$.

\textbf{Assumption A1 (Identification).}
The parameter $\psi_0$ is identifiable under consistency,
conditional ignorability, and positivity.

\textbf{Assumption A2 (Influence Function Representation).}
The estimator $\hat\psi_n$ admits the asymptotic linear expansion
\[
\sqrt{n}(\hat\psi_n - \psi_0)
=
\frac{1}{\sqrt{n}}\sum_{i=1}^n \Phi(O_i) + o_p(1),
\]
where $\Phi$ is the vector of influence functions of $\psi$ under $P_0$.

\textbf{Assumption A3 (Moment Conditions).}
The influence function satisfies
\[
\mathbf{E}_{P_0}[\Phi(O_i)] = 0,
\] and that $\Sigma = \mathbf{E}_{P_0}[\Phi(O_i)\Phi(O_i)^T]$ is  finite and positive definite.

Then,
\[
\sqrt{n}(\hat\psi_n - \psi_0)
\xrightarrow{d}
\mathcal{N}(0, \Sigma).
\]
Moreover, the variance estimator
\[
\hat\Sigma_n = \frac{1}{n}\sum_{i=1}^n\hat\Phi(O_i)\hat\Phi(O_i)^T\xrightarrow{p} \Sigma,
\]
where $\hat{\Phi}(O_i)$ is equal to $\Phi(O_i)$ with the unknown parameter $\psi_0$ replaced by $\hat{\psi}_n$.
\end{theorem}

The influence functions under the three proposed methods in Sections 2.1-2.3 take the following forms 
\begin{equation} IF_i(\text{IPTW}, Z=1, k)=\phi_{i1k}(\text{IPTW})\approx w_iZ_iI(Y_i=k)+q_{1k}^TU_{i,ps}(\beta)-p_{1k}, \end{equation}
\begin{equation} IF_i(\text{G-Form},Z=a, k)=\phi_{iak}(\text{G-Form})\approx m_{iak}+l_{ak}^TU_{i,po}(\Theta)-p_{ak}, \end{equation}
\begin{equation} 
\begin{aligned}
\text{and }IF_i(\text{Doubly Robust},Z=1, k)=\phi_{i1k}(\text{DR})
&\approx \{w_iZ_iI(Y_i=k)-(w_iZ_i-1)m_{i1k}-p_{1k}\}\\
&+U_{i,ps}(\beta)\times\frac{1}{n}\sum_{i=1}^nw_i'Z_i[I(Y_i=k)-m_{i1k}]\\
&-U_{i,po}(\Theta)\times\frac{1}{n}\sum_{i=1}^nm_{i1k}'[w_iz_i-1].
\end{aligned}
\end{equation}

Intuitively speaking, $w_i$ is the weight inversely calculated from $\pi_i$, $U_{i,ps}(\beta)$ is the individual influence function for $\beta$ under the propensity score model, $U_{i,po}(\Theta)$ is the individual influence function for $\Theta$ under the potential outcome model, and $q_{1k}, l_{ak}$ are vectors of partial derivatives under the corresponding models. The IFs for $Z=0$ in IPTW and DR have similar expressions by replacing the occurrence of $Z_i$ with $1-Z_i$ and $m_{i1k}$ with $m_{i0k}$. More details can be found in the Appendix.\\

By combining all estimated Influence Function values of an individual into a vector, e.g. $\hat\Phi(O_i)=[\hat\phi_{i11}, \hat\phi_{i12}, \cdots, \hat\phi_{i1(K-1)}, \hat\phi_{i01}, \cdots, \hat\phi_{i0(K-1)}]$ (note that we only need to estimate $K-1$ such influence values for the Delta method in the next step), and combining $n$ such Influence vectors of length $2K-2$ into a matrix, we have a matrix with dimension $2K-2$ by $n$. Consequently, we can use the outer product of this matrix with itself divided by $n$, a square matrix of dimension $2K-2$, as the estimate of the covariance matrix, where the diagonal blocks are the variance matrices of $P(Y^1)$ and $P(Y^0)$, and the off-diagonal blocks are the covariance between the counterfactuals.\\

Finally, suppose by all previous steps that we can estimate a vector of probabilities $[\hat p_{11},\hat p_{12},\cdots$, $\hat p_{1(k-1)},\hat p_{01},\cdots,\hat p_{0(k-1)}]$ and corresponding covariance $\hat\Sigma_n$, we can derive the partial derivatives of DOOR probability with respect to each cell probability as $J$, we have $\hat P(Y^1)^TA \hat P(Y^0)=\hat D\sim N(D, J^T\hat\Sigma_nJ)$ by the Delta Method. Detailed procedure for estimating the partial derivatives can be found in the Appendix. Note that the Delta Method here is independent of the causal inference method used before and can be applied as long as $\Sigma$ can be estimated.\\

In implementing the methods described above, we rely on the {\it lava} R package \citep{c16} to conduct the proportional odds regression (function ordreg) and retrieve influence functions from the parametric propensity score model and the potential outcome model (function IC).\\

\section{Results}
\subsection{Simulation}

Without loss of generality, we consider four covariates, each with mean 0 and variance 1, and a pairwise correlation coefficient of 0.5. Note that the covariance structure is not of importance here, but the idea is that we want to generate $X$'s that are correlated, which is typical in biomedical study, e.g., Body Mass Index and Hemoglobin A1c. The first two covariates $X_1, X_2$ will be set as continuous, and the next two $X_3, X_4$ will be dichotomized at zero to be binary. We have $\mathbf{X} = (X_1,X_2,X_3,X_4)$.\\

We assume that the probability of receiving treatment follows some true propensity score model, $\text{logit}\{P(Z_i=1|X)\}=\beta_0+\beta^TX$, where $\beta_0=-0.4, \beta=(0.15, -0.3, 0.2, -0.25)$. Following the propensity score setup, we define $e=\exp(\beta_0+\beta^T X), \pi=1/(1+e)$. Treatment $Z$ follows a Bernoulli distribution with a success probability $\pi$. In this particular setup, we have the prevalence of treatment at around 0.4.\\

The outcome $Y$ with $K=4$ levels is generated from a separate outcome model: $\text{log}(P(Y\leq k\mid\mathbf{X})/P(Y>k\mid\mathbf{X})) = \alpha_k + \mathbf{X}^T\gamma+\Delta Z, k=1,\cdots,K-1$, where $(\alpha_1,\alpha_2,\alpha_3)=(1, 0.5, -0.5)$ contains level-specific intercepts, $\gamma=(0.8,-0.4,0.6,-0.3)$ contains coefficients for covariates, and $\Delta$ is the treatment effect. We invert the log-odds to retrieve the vector $P(Y)$ and determine the observed outcome by sampling one category from the $K$ categories with the corresponding probabilities. We approximate the true DOOR probabilities through a million Monte Carlo simulations. Note that for this Monte Carlo estimation to calculate the true DOOR probabilities, we need to construct counterfactuals, i.e., $X$ can be all random, but the treatment must be fixed at $0$ and $1$, so we can get the true DOOR probability.\\

We consider four different scenarios with respect to our proposed methods: both models are correctly specified, only the Propensity Score model is correctly specified, only the Potential Outcome model is correctly specified, and both models are incorrectly specified. A correctly specified model will include all four covariates, while an incorrectly specified model only contains $X_1$ and $X_3$ but misses the other two variables. In each scenario, we estimate the DOOR probability from the three methods described above. Table 1 reports empirical bias, empirical standard deviation of parameter estimates, empirical average of standard error estimates, empirical coverage probability of the 95\% confidence intervals under two different sample sizes, all based on 10000 replicates. In this setting, we set $\Delta=0.4$, resulting in the true DOOR probability of 0.552. We compare the results of our three proposed methods with a crude estimator that does not adjust for any covariates.\\

\newpage

\begin{table}[h]
\caption{Summary statistics of the estimators with $N$ participants, based on 10000 replicates.}
\begin{center}
\begin{tabular}{lcccccccc} \hline
Method & Bias & SE$^a$ & SEE$^b$ & CP$^c$ & Bias & SE & SEE & CP\\
 & & $N=500$ & & & & $N=1000$ & &\\
Crude & 0.013 & 0.024 & 0.024 & 0.922 & 0.013 & 0.017 & 0.017 & 0.879\\
IPTW & & & & & & & &\\
Correct$^d$ & -0.001& 0.023& 0.024& 0.960 & -0.001& 0.017& 0.017& 0.958\\
Incorrect$^e$ & 0.014& 0.023& 0.024& 0.912 & 0.014& 0.016& 0.017& 0.867\\
G-Formula & & & & & & & &\\
Correct & -0.001& 0.023& 0.023& 0.946 & -0.001& 0.016& 0.016& 0.942\\
Incorrect & 0.014& 0.023& 0.022& 0.890 & 0.014& 0.016& 0.016& 0.843\\
Doubly Robust & & & & & & & &\\
Both Correct & -0.001& 0.023& 0.024& 0.961 & -0.001& 0.017& 0.017& 0.957\\
PS Correct & -0.001& 0.023& 0.024& 0.960 & -0.001& 0.017& 0.017& 0.958\\
PO Correct & -0.001& 0.023& 0.024& 0.957 & -0.001& 0.017& 0.017& 0.955\\
Both Incorrect & 0.014& 0.023& 0.024& 0.912 & 0.014& 0.016& 0.017& 0.867\\
\hline
\vspace{-8mm}
\label{t1}
\end{tabular}
\end{center}
\end{table}
{\footnotesize{$^a$ SE: empirical standard deviation of parameter estimates based on 10000 replicates; $^b$ SEE: empirical average of standard error estimates based on 10000 replicates; $^c$ CP: empirical coverage probability of the 95\% confidence intervals based on 10000 replicates; $^d$Correct: model is correctly specified, i.e., all four covariates are included; $^e$Incorrect: model is incorrectly specified, i.e., only $X_1$ and $X_3$ are included}}\\

\newpage

We can make a few observations from Table 1: 1. When all models are correctly specified, IPTW/G-Formula (singly robust) and Doubly Robust methods have negligible bias and near-nominal coverage rates, where the IPTW and Doubly Robust methods slightly overcover and the G-Formula slightly undercovers; 2. When one model is correctly specified, IPTW or G-Formula method becomes no better than the crude method if the corresponding model is misspecified. However, the Doubly Robust method has another layer of protection, as it only requires one model to be correctly specified to ensure unbiased estimates, and guarantees low bias and nominal coverage rate; 3. When no models are correctly specified, the Doubly Robust method becomes similar to other methods. The increase in sample size decreases the standard error estimates, but the bias persists, reducing the coverage rate even further. Specifically, these covariate adjustment methods for DOOR analysis mainly focus on debiasing, rather than reducing uncertainty; 4. The proposed methods, especially the Doubly Robust method, are slightly conservative, since the empirical coverage rate of 95\% CI is around 96\% when at least one model is correct, reflecting the trade-off between efficiency and robustness.\\

Additionally, we conducted simulations with different values of $\Delta=(0,0.05,0.1,\cdots,0.4)$, resulting in DOOR probabilities from 0.5 to 0.55, and compared the empirical powers under different methods. The results are presented in Table 2. From Table 2, we have findings consistent with the results in Table 1: 1. Conservative Confidence Intervals from the IPTW and DR methods result in conservative Type I error control, whereas the G-formula method has slightly better results when the model is correctly specified. 2. The crude method and misspecified models seem to have a higher power curve, but the cause behind the scene is that they have highly inflated type I error rates. They estimate the DOOR probability with a large bias, and thus it is always easier to reject the null, compared to correct models. 3. A larger sample size or effect size will increase the power to reject the null.

\begin{table}[h]
\caption{Empirical powers based on 10000 replicates. For each DOOR probability, we test against the null hypothesis  $D=0.5$. C indicates correctly specified model, while IC indicates misspecified models.}
\begin{center}
\begin{tabular}{lccccccccc} \hline
$D$ & 0.500 & 0.507 & 0.514 & 0.521 & 0.527 & 0.533 & 0.540 & 0.546& 0.552\\
 & & & &$N=500$ & & & &\\
Crude & 0.088 & 0.134 & 0.201 & 0.270 & 0.377 & 0.485 & 0.570 & 0.678 & 0.755\\
IPTW C & 0.037& 0.044& 0.072& 0.117& 0.183& 0.267& 0.354&0.461&0.568\\
IPTW IC & 0.088& 0.135& 0.214& 0.293& 0.400& 0.514& 0.606& 0.718&0.791\\
G-Form C & 0.050& 0.060& 0.094& 0.145& 0.222& 0.313& 0.411& 0.519&0.623\\
G-Form IC & 0.106& 0.160& 0.246& 0.329& 0.445& 0.558& 0.647&0.750&0.821\\
DR Both C & 0.037& 0.043& 0.072& 0.118& 0.183& 0.266& 0.358& 0.461&0.568\\
DR PS C & 0.037& 0.044& 0.073& 0.118& 0.183& 0.268& 0.357& 0.462&0.567\\
DR PO C & 0.038& 0.046& 0.076& 0.121& 0.191& 0.275& 0.368&0.471&0.579\\
DR Both IC & 0.088& 0.135& 0.212& 0.293& 0.400& 0.513& 0.606& 0.717&0.791\\
 & & & &$N=1000$ & & & &\\
Crude & 0.130 & 0.217 & 0.343 & 0.473 & 0.635 & 0.754 & 0.861 & 0.923&0.964\\
IPTW C & 0.039& 0.055& 0.108& 0.184& 0.318& 0.471& 0.618&0.757&0.854\\
IPTW IC & 0.137& 0.226& 0.364& 0.507& 0.676& 0.795& 0.888& 0.946&0.975\\
G-Form C & 0.052& 0.070& 0.135& 0.228& 0.373& 0.528& 0.671& 0.799&0.884\\
G-Form IC & 0.163& 0.261& 0.411& 0.553& 0.715& 0.826& 0.906&0.955&0.980\\
DR Both C & 0.039& 0.054& 0.107& 0.185& 0.317& 0.470& 0.618& 0.758&0.855\\
DR PS C & 0.039& 0.055& 0.107& 0.185& 0.318& 0.471& 0.618& 0.757&0.854\\
DR PO C & 0.042& 0.057& 0.113& 0.192& 0.329& 0.483& 0.630&0.766&0.862\\
DR Both IC & 0.137& 0.226& 0.365& 0.507& 0.676& 0.795& 0.886& 0.947&0.976\\
\hline
\vspace{-8mm}
\label{t2}
\end{tabular}
\end{center}
\end{table}

\newpage

\subsection{Real-world application}

Multi-Drug Resistant Organism (MDRO) network is a study within the Antibacterial Resistant Leadership Group (ARLG) that provides observational data to aid in the design of randomized clinical trials on therapeutics and diagnostics for MDRO infections. Previous works have studied clinical outcomes and characteristics of Carbapenem-resistant \textit{Acinetobacter Baumannii} (SNAP \cite{c30}), \textit{Pseudomonas aeruginosa} (POP, \cite{c29}, and \textit{Enterobacterales} (CRACKLE, \cite{c40}). In this analysis, we combine the data from the three studies to compare the benefit:risk profile of Mono-drug vs. Combination-drug treatment during the first two days after patient sample culture collection. The final analysis cohort contains 2333 patients, where 1714 receive mono-therapy and 619 receive combination-therapy. For covariates, we consider geographical location, study/bacteria type, Pitt bacteremia score (the higher the worse), Age-adjusted Charlson Index (the higher the worse), and infection site.\\

The DOOR variable is defined as the following (from most desirable to least desirable): 1. Alive without any of the three qualifying events 2. Alive with one of the three qualifying events 3. Alive with two or three of the three qualifying events 4. Death. The three qualifying events include absence of clinical response, infectious complications, and non-fatal serious adverse events (SAE). From Table 3 we can see the difference in baseline patient characteristics between the two treatments: Patients who received mono-therapy have a higher chance to be in the most desirable status possible, have a lower risk indicated by Pitt score, and have a smaller percentage with blood infection, which is considered as a more serious infection type. A crude estimation of the DOOR probability might lead to biased results.\\

\begin{table}[h]
\centering
\caption{Baseline Patient Characteristics by Treatment Type. P-values were calculated using Pearson’s Chi-square test for categorical variables and Kruskal–Wallis test for continuous variables.}
\vspace{5mm}
\begin{tabular}{llll}
\hline
Characteristics/Treatment & Mono-therapy & Combination-therapy & p-value \\ 
& $(n=1714)$ & $(n=619)$ & \\
\hline
DOOR, $n(\%)$ & & & $<0.001$ \\ 
\quad 1& 637(37) & 166(27) & \\ 
\quad 2& 412(24) & 156(25) & \\ 
\quad 3& 315(18) & 128(21) & \\ 
\quad 4& 350(20) & 169(27) & \\ 
Country, $n(\%)$ & & & 0.025 \\ 
\quad USA& 874(51) & 287(46) & \\ 
\quad Asia-Pacific& 53(3.1) & 11(1.8) & \\ 
\quad China& 525(31) & 202(33) & \\ 
\quad Middle East& 73(4.3) & 40(6.5) & \\ 
\quad South America& 189(11) & 79(13) & \\ 
Study/Bacteria, $n(\%)$ & & & $<0.001$ \\ 
\quad CRACLE(CRE) & 979(57) & 375(61) & \\ 
\quad POP(CRPA) & 470(27) & 119(19) & \\ 
\quad SNAP(CRAB) & 265(15) & 125(20) & \\ 
Infection, $n(\%)$ & & & $<0.001$\\ 
\quad Blood & 463(27) & 266(43) & \\ 
\quad Respiratory & 722(42) & 250(40) & \\ 
\quad Urine & 186(11) & 40(6.5) & \\ 
\quad Other & 343(20) & 63(10) & \\ 
Pitt Score, Median IQR) & 3.0(1.0-6.0) & 4.0(2.0-6.0) & $<0.001$\\ 
Charlson index, Median(IQR) & 4.0(2.0-6.0) & 4.0(2.0-6.0) & 0.16\\ \hline
\end{tabular}
\label{t3}
\end{table}

We consider the same list of covariates described above as confounders for the propensity model and the outcome model. In Table 4, there are a few observations we can make: 1. The crude analysis, which does not adjust for any covariates, yields a smaller DOOR probability estimate, indicating that Mono-therapy could be more ideal for patients, but this estimate, combined with the patient differences in Table 3, could be biased towards Mono-therapy; 2. The three covariate-adjustment methods proposed yield a point estimate that favors more toward the null hypothesis compared to the crude analysis; 3. The interpretation of this analysis indicates that a patient receiving combination-therapy has a probability of around 0.47 to be in a more desirable status than the same patient receiving mono-therapy; 4. The upper bound of the confidence intervals is below 0.5, which indicates that mono-therapy has a statistically significant advantage over combination-therapy conditional on all the confounders considered. 5. All three methods give similar estimates of the treatment effect, which could be due to the fact that the IPTW and G-Formula methods already include important confounders that could bias our analysis; in a subsequent analysis not shown here, we identify the Pitt bacteremia score as a potential driver of the confounding effect.\\

Since DOOR is a composite outcome, a fundamental tenet of analyses of composites is the incorporation of the analysis of each of the components to ensure a thorough understanding of the manifestation of effects, and we display the sequentially dichotomized DOOR probability in Figure 1. For simplicity, we only consider the doubly robust method for covariate adjustment here. The sequential analysis demonstrates that the advantage of Mono-therapy lies more in preventing mortality. All comparisons made are not meant to be the final assessment of mono-therapy vs. combination-therapy, but the results here could become the basis for future trial design to answer this research question.

\begin{table}[h]
\centering
\caption{DOOR probability estimate under different methods for the MDRO study}
\vspace{5mm}
\begin{tabular}{lll}
\hline
Method & Estimate & 95\% Confidence Interval\\ 
Crude & 0.436 & (0.410, 0.461) \\ 
IPTW & 0.470 & (0.444, 0.495) \\ 
G-Formula & 0.470 & (0.444, 0.495) \\ 
Doubly Robust & 0.471 & (0.446, 0.496) \\ \hline
\end{tabular}
\label{t4}
\end{table}

\begin{figure}[h]
\begin{center}
\includegraphics[width=6in]{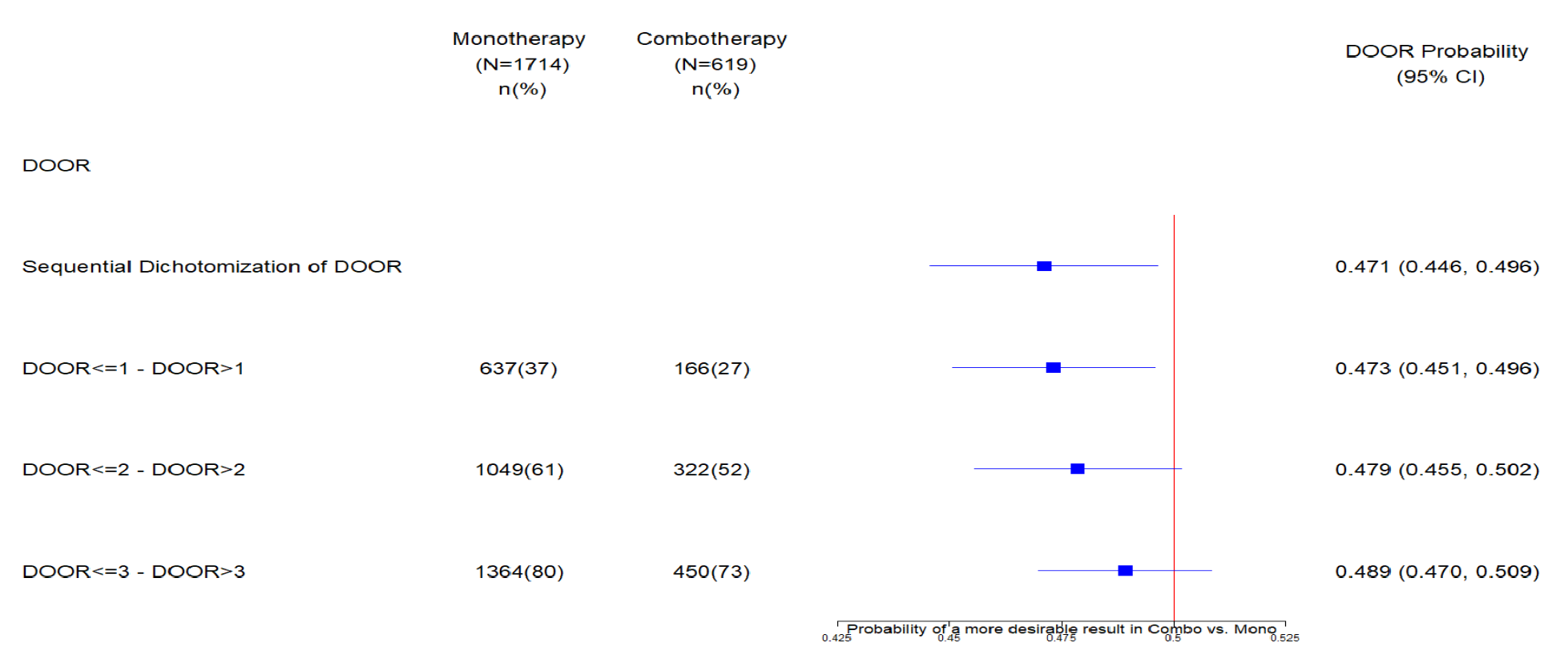}
\end{center}
\caption{Forest plot of Sequentially Dichotomized DOOR Probability Estimate with the Doubly Robust method}
\label{fig1}
\end{figure}

\section{Discussion}

In this manuscript, we present a framework for estimating the DOOR probability with covariate adjustment. Our approach in deriving the IPTW, the G-Formula, and the doubly robust/augmented IPTW estimators is closely related to the procedure discussed in \cite{c18}, but we do not rely on the contrast/kernel function in U-statistics, and instead derive the asymptotic properties based on influence functions and the Delta method. From the simulations, we can observe that whether the model is correctly or incorrectly specified, the standard error estimates are similar across methods, and the discrepancy in coverage rate is due to the bias caused by misspecified models. A correctly specified G-Formula model based on maximum likelihood estimation can be more efficient in terms of coverage rate and type I error control, but the doubly robust method offers additional protection against model misspecifications (as in the discussion of \cite{c17}). The empirical coverage rates of the 95\% confidence interval of the doubly robust method when at least one model is correct, are slightly larger than 95 percent, and the empirical type I error rate  is smaller than 5 percent, indicating a conservative method. As shown in the Appendix, the inference procedure undergoes three Taylor's expansion/approximation (influence functions of the parametric model(s), influence of functions for the DOOR distribution, the Delta method to estimate the variance of DOOR probability), and there might be a slight loss in the accuracy of the standard error estimation. The doubly robust estimator could remain valid even when one model is misspecified, so our method can be considered as a valuable initial step in covariate adjustment for DOOR analysis. Our proposed asymptotic variance estimator is based on influence functions, and can be considered as a sandwich estimator where the ``bread" is the derivative of the estimating equation, and the ``meat" is the empirical variance of the influence function. Alternatively, we can consider the bootstrap method by sampling with replacement a large number of data sets. We then use the empirical standard deviation of the Bootstrap estimates of the DOOR probability to estimate the standard error. Logistics regression for the propensity model and proportional odds regression for the outcome model are our major focus here, but alternative methods such as the Probit regression can be plugged into our formula/theorem as long as we have the influence functions of the regression parameters.\\

The current G-Formula approach constructs a single outcome model with pooled data, which assumes the same functional form across treatment arms and can be efficient (lower variance). We can also fit separate models for data within the treatment arm and data within the control arm, which allows flexibility in model specification, and can be safer when strong effect modification is present. Alternatively, we can include an interaction term between treatment and confounders in the pooled model. Regardless of the outcome model(s) we fit, the steps of estimating the DOOR probability and making inference should be similar, and a comparison of the previous options could be a future direction.\\

Since DOOR methodology can be considered as one example of Generalized Pairwise Comparison (GPC), we can easily extend the doubly robust approaches discussed here to other estimands such as Net Treatment Benefit (NTB, \cite{c10}), Win Ratio \citep{c11}, Win Odds \citep{c1} etc. There are two main advantages of developing such methods in the DOOR framework versus other estimands: 1. The first step of DOOR requires the construction of an ordinal variable, which can be readily used for regression models, while other methods hierarchically compare multiple outcomes; 2. DOOR analysis could avoid issues associated with the relative risk metrics\cite{c32, c34}, particularly in multi-outcome problems, as recommended in \cite{c33}.\\

As discussed in \cite{c19}, DOOR probability is a U-statistic that undergoes a nonlinear transformation of counterfactual probabilities, so it will have different interpretations at the population-level and the individual-level. In this manuscript, we consider the population-level approach of averaging counterfactual probabilities by intervention arm first, and calculating the population-level DOOR probability, which syncs with our proposed IPTW and Doubly Robust method. The alternative way will be to calculate the individual DOOR probabilities based on the counterfactual probabilities and then take the average, but this approach requires an outcome model and its correct specification. The individual-level DOOR probability estimator takes the form of the average of functions of model parameters, and its variance estimation will be a future topic.\\

We envision the primary usage of the proposed method to be in observational studies, where the doubly robust mechanism allows the flexibility of specifying different models for the treatment and the outcome. It is reasonable to assume that different lists of confounders would contribute to how a patient receives treatment and what the outcome of this patient will be. Even in clinical trials, the introduction of covariate adjustment methods could correct for stratified or blocked randomization \citep{c5}, and improve precision \citep{c20,c21}. Regression models are sometimes required in the statistical analysis plan for regulatory submission \citep{c6,c37,c38}, and our methods could also be extended in that direction.\\

Some future directions could be considered: 1. A better understanding of the proposed methods. Since we only consider a single scenario where both propensity model and outcome model are generated by the same four covariates, we can evaluate the performance by introducing interaction terms, non-linear terms, or unobserved variables. 2. We can incorporate other techniques in causal inference (e.g., overlapping weights\citep{c26}, TMLE\citep{c27}). 3. We can extend the covariate adjustment methods discussed here to other relevant DOOR analysis (e.g., longitudinal analysis\citep{c24}, or survival analysis\citep{c23, c25}).\\

\bigskip
\begin{center}
{\large\bf SUPPLEMENTARY MATERIAL}

\end{center}

\section*{Appendix}

We outline the proofs of the asymptotic results of the DOOR probability estimators using the IPTW, G-Fomrula, and doubly-robust methods presented in the Methods section.

\subsection*{IPTW Equation (2)}
Define $w_i=1/\pi_i=1+\exp(-\beta^TX_i)$ when $Z=1$, and $=1/(1-\pi_i)=1+\exp(\beta^TX_i)$ when $Z=0$. Consequently, we have the first derivative of $w_i$ with respect to $\beta$ as $w'_i=-X_i\exp(-\beta^TX_i)$ when $Z=1$, and $X_i\exp(\beta^TX_i)$ when $Z=0$. The IPTW estimator of counterfactual probabilities can be expressed as $\hat p_{1k}=\frac{1}{n}\sum_{i=1}^n\hat w_iZ_iI(Y_i=k), \hat p_{0k}=\frac{1}{n}\sum_{i=1}^n\hat w_i(1-Z_i)I(Y_i=k)$.\\

Applying the Taylor expansion to $w_i$ as a function of $\beta$, we have $\sqrt{n}(\hat w_i-w_i)=\sqrt{n}(\hat\beta-\beta)w_i'+op(1)$. Then we have $\hat w_i\approx w_i+(\hat\beta-\beta)w_i'$. By subtracting $p_{1k}$ on both sides, we have $\hat p_{1k}-p_{1k}=\frac{1}{n}\sum_{i=1}^n\hat w_iZ_iI(Y_i=k)-p_{1k}$, and $\sqrt{n}(\hat p_{1k}-p_{1k})=\frac{1}{\sqrt{n}}\sum_{i=1}^n\{[w_iZ_iI(Y_i=k)-p_{1k}]+[(\hat\beta-\beta)w_i'Z_iI(Y_i=k)]\}$ by replacing $\hat w_i$ with the Taylor expansion. The previous expression is thus separated into two sums, $\frac{1}{\sqrt{n}}\sum_{i=1}^n[w_iZ_iI(Y_i=k)-p_{1k}]+\sqrt{n}(\hat\beta-\beta)[\frac{1}{n}\sum_{i=1}^n w_i'Z_iI(Y_i=k)]$. We can further write $\sqrt{n}(\hat\beta-\beta)=\frac{1}{\sqrt{n}}\sum_{i=1}^nU_{i,ps}(\beta)+op(1)$, and $\frac{1}{n}\sum_{i=1}^n w_i'Z_iI(Y_i=k)=q_{1k}$, the same value that can be empirically estimated for all individuals. Note that $q_{1k}$ can be seen as an average of the derivatives with respect to the parameters in the propensity score model, matching the dimension of $U_{ps}(\beta)$.\\

Therefore, we have $\sqrt{n}(\hat p_{1k}-p_{1k})=\frac{1}{\sqrt{n}}\sum_{i=1}^n[w_iZ_iI(Y_i=k)-p_{1k}]+\frac{1}{\sqrt{n}}\sum_{i=1}^nq^T_{1k}U_{i,ps}(\beta)+op(1)$, which satisfies the form of an influence function and matches the expression in Equation 2. By the same fashion, we can write $\sqrt{n}(\hat p_{0k}-p_{0k})=\frac{1}{\sqrt{n}}\sum_{i=1}^n[w_i(1-Z_i)I(Y_i=k)-p_{0k}]+\frac{1}{\sqrt{n}}\sum_{i=1}^nq^T_{0k}U_{i,ps}(\beta)+op(1)$.

\subsection*{G-Formula Equation (3)}

The derivations here are similar to those in the previous section. $\sqrt{n}(\hat p_{1k}-p_{1k})=\frac{1}{\sqrt{n}}\sum_{i=1}^n\hat m_{i1k}-p_{1k}, \sqrt{n}(\hat m_{i1k}-m_{i1k})=\sqrt{n}(\hat \Theta-\Theta)m'_{i1k}+op(1)$. By replacing $\hat m_{i1k}$ with the remaining terms after the Taylor expansion, we have $\sqrt{n}(\hat p_{1k}-p_{1k})=\frac{1}{\sqrt{n}}\sum_{i=1}^n(\frac{1}{n}\sum_{i=1}^nm_{i1k}')^TU_{i,po}(\Theta)+m_{i1k}-p_{1k}=\frac{1}{\sqrt{n}}\sum_{i=1}^n l_{1k}^TU_{i,po}(\Theta)+m_{i1k}-p_{1k}$.\\

There are two remarks that we would like to state in the derivatives of $m_{i1k}$: 1. We need to pay attention to the parameterization of ordinal regression. Classical parameterization has $P(Y_i\leq k)=\frac{exp(\alpha_k+\beta^TX_i)}{1+exp(\alpha_k+\beta^TX_i)}$, but the ordreg function we use from the {\it lava} R package utilizes $P(Y_i=1 )=\frac{exp(\alpha_1+\beta^TX_i)}{1+exp(\alpha_1+\beta^TX_i)}, P(Y_i\leq k, k\geq 2)=\frac{exp(\alpha_1+\sum_{j=2}^kexp(\alpha_j)+\beta^TX_i)}{1+exp(\alpha_1+\sum_{j=2}^kexp(\alpha_j)+\beta^TX_i)}$. (This parameterization could guarantee monotonicity in the intercept estimates) 2. $m_{i1k}=P(Y_i\leq k)-P(Y_i\leq k-1)$, and the partial derivatives with respect to each parameter can be started from here.

\subsection*{Doubly Robust Equation (4)}

With all the definitions and notation above, we can easily combine the results to derive the influence functions for the doubly robust method:

\begin{equation} 
\begin{aligned}
\hat p_{1k}
&=\frac{1}{n}\sum_{i=1}^nw_iZ_iI(Y_i=k)-(w_iZ_i-1)\hat m_{i1k},\\
\sqrt{n}(\hat p_{1k}-p_{1k})&=\frac{1}{\sqrt{n}}\sum_{i=1}^n(w_i+w'_i(\hat \beta-\beta))Z_iI(Y_i=k)\\
&-((w_i+w_i'(\hat \beta-\beta))Z_i-1)(m_{i1k}+m'_{i1k}(\hat\Theta-\Theta))-p_{1k},\\
&=\frac{1}{\sqrt{n}}\sum_{i=1}^nw_iZ_iI(Y_i=k)+(\hat\beta-\beta)w_i'Z_iI(Y_i=k)-p_{1k}\\
&-w_iZ_iM_{i1k}-(\hat\Theta-\Theta)w_iZ_im'_{i1k}-(\hat\beta-\beta)w_i'Z_im_{i1k}+m_{i1k}+(\hat\Theta-\Theta)m_{i1k}'\\
&-(\hat\beta-\beta)(\hat\Theta-\Theta)w_i'Z_im_{i1k},\\
&=\frac{1}{\sqrt{n}}\sum_{i=1}^n\{w_iZ_iI(Y_i=k)-(w_iZ_i-1)m_{i1k}-p_{1k}\} \hfill \text{ (term 1) }\\
&+\{(\hat\beta-\beta)[w_i'Z_iI(Y_i=k)-w'_iZ_im_{i1k}]\}\hfill \text{ (term 2) }\\
&-\{(\hat\Theta-\Theta)[(w_iZ_i-1)m'_{i1k}]\} \hfill \text{ (term 3) }\\
&-(\hat\beta-\beta)(\hat\Theta-\Theta)w_i'Z_im_{i1k} \hfill \text{ (term 4).}
\end{aligned}
\end{equation}

Intuitively, we can interpret term 1 as $p_{i1k}-E(p_{i1k})$, term 2 as the contribution from IPTW, term 3 as the contribution from G-computation, and term 4 as negligible. Again, parameters from the propensity score model and the potential outcome model can be replaced by the corresponding influence functions expression in the equation.

\subsection*{Delta Method for DOOR probability}

We can express the DOOR probability as $D=\frac{1}{2}p_{11}p_{01}+[\frac{1}{2}p_{12}p_{02}+p_{12}p_{01}]+[\frac{1}{2}p_{13}p_{03}+p_{13}(p_{01}+p_{02})]+\cdots+[\frac{1}{2}(1-p_{11}-p_{12}-\cdots-p_{1(K-1)})(1-p_{01}-p_{02}-\cdots-p_{0(K-1)})+(1-p_{11}-p_{12}-\cdots-p_{1(K-1)})(p_{01}+p_{02}+\cdots+p_{0(K-1)})]$. Note that we replace the last probability with the difference between 1 and the previous cumulative probability. By taking partial derivatives with respect to $p_{1k}$, we have

\begin{align*}
    \frac{\partial D}{\partial p_{11}}&=-\frac{1}{2}-\frac{1}{2}(p_{02}+\cdots+p_{0(K-1)})\\
     \frac{\partial D}{\partial p_{12}}&=-\frac{1}{2}+\frac{1}{2}p_{01}-\frac{1}{2}(p_{03}+\cdots+p_{0(K-1)})\\
     \frac{\partial D}{\partial p_{1(K-1)}}&=-\frac{1}{2}+\frac{1}{2}(p_{01}+p_{02}+\cdots+p_{0(K-2)})\\
\end{align*}

Following this derivation, we can write in a generic form $\frac{\partial D}{\partial p_{1j}}=\frac{p_{0j}-p_{0K}}{2}-\sum_{i=1}^{K-1}p_{0i}, \frac{\partial D}{\partial p_{0j}}=\frac{p_{1j}-p_{1K}}{2}-\sum_{i=1}^jp_{1i}, j=1,\cdots,K-1$. By saving these derivatives as a vector, we have $J$ as described in the Methods section.

\bibliographystyle{agsm}

\bibliography{bibliography}

\end{document}